\documentstyle[prl,aps,epsf,multicol]{revtex}

\makeatletter
\def\preprint#1{%
\def\@preprint{\noindent\hfill\hbox{#1}\vskip 10pt}%
}
\makeatother

\begin{document}
\preprint{ITP-UH-21/96, OUTP-9632S}
\title{Integrable impurity in the supersymmetric $t$-$J$ model}
\vspace{1.5 em}

\author{Gerald Bed\"urftig$^1$, Fabian H.\,L.\,E\ss{}ler$^2$, and
	Holger Frahm$^1$}
\address{$^1$Institut f\"ur Theoretische Physik, Universit\"at Hannover,
	 D-30167~Hannover, Germany}
\address{$^2$Department of Physics, Theoretical Physics,
	Oxford University\\ 1 Keble Road, Oxford OX1 3NP, Great Britain}
\date{September 1996}
\maketitle
\begin{abstract}
An impurity coupling to both spin and charge degrees of freedom is added to
a periodic $t$-$J$ chain such that its interaction with the bulk can be
varied continuously without losing integrability.  Ground state properties,
impurity contributions to the susceptibilities and low temperature
specific heat are studied as well as transport properties. The
impurity phase--shifts are calculated to establish the existence of an
impurity bound state in the holon sector.
\end{abstract}

\begin{multicols}{2}
\narrowtext
Quantum fluctuations are known to play an important role in the physics of
low dimensional strongly correlated electron systems: the low temperature
properties of such systems in one spatial dimension have to be described in
terms of a Luttinger liquid rather than a Fermi liquid.  {}From an
experimental point of view the transport properties of these systems in the
presence of boundaries and potential scatterers are of particular interest.
Recently several attempts have been made to describe such a situation:
Using renormalization group techniques the transport properties of a 1D
interacting electron gas in the presence of a potential barrier have been
studied by Kane and Fisher \cite{kafi:92}.  Their surprising findings
triggered further work using different techniques like boundary conformal
field theory \cite{card:89} and an exact solution by means of a mapping to
the boundary sine-Gordon model \cite{Fendley,tsve:95a}.  In particular the
low temperature properties of magnetic (Kondo) impurities in a Luttinger
liquid \cite{Affleck,FrojdhJ} have been investigated in
great detail.  In the present work we will investigate the effects of a
particular type of potential impurity in a Luttinger liquid (where both
spin-and charge degrees of freedom are gapless) by means of an exact
solution through the Quantum Inverse Scattering Method (QISM)
\cite{vladb}. 

Attempts to study effects due to the presence of impurities in many-body
quantum system in the framework of integrable models have a long successful
history \cite{anjo:84,LeeSchl,Vega,bares:95,KondoBA}.  As far as lattice
models are concerned the underlying principle in these exact solutions is
the fact that the QISM allows for the introduction of certain
``inhomogeneities'' into vertex models without spoiling integrability.
The {\em local} vertices---so called ${\cal L}$-operators---are objects
depending on a complex valued spectral parameter acting on an auxiliary
matrix space in addition to the quantum space of the model.  They are
solutions of a Yang-Baxter equation with an ${\cal R}$ matrix which itself
acts on two copies of the matrix space and depends on the difference of the
corresponding spectral parameters only.  This allows to to build families
of vertex models with site-dependent shifts of the spectral parameters and
even different quantum spaces on different sites. The first fact has been
widely used in solving models for particles with an internal degree of
freedom by means of the nested Bethe Ansatz \cite{yang:67}.  The second
approach has been first applied by Andrei and Johannesson to study the
properties of an $S={1\over2}$ Heisenberg chain with an additional site
carrying spin $S$ \cite{anjo:84} (see also \cite{LeeSchl}).

In this letter we study the properties of the supersymmetric $t$--$J$ model
with one vertex replaced by an ${\cal L}$ operator acting on a
four--dimensional quantum space. This preserves the $gl(2|1)$ supersymmetry
of the model but at the same time lifts the restriction of no double
occupancy present in the $t$--$J$ model at the impurity site. The existence
of a free parameter in the four--dimensional representations of the
superalgebra \cite{brax:94} allows to tune the coupling of the impurity to
the host chain.  As will be shown below, the present model allows to study
some aspects of a more general situation than the ones mentioned above: the
impurity introduced here couples to {\sl both} spin- and charge degrees of
freedom of the bulk Luttinger liquid. The extension of our calculation to
the case of many impurities is straightforward.

\begin{figure}
\noindent
\epsfxsize=0.45\textwidth
\epsfbox{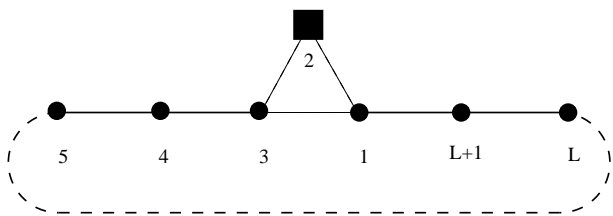}
\caption{\label{fig:im.ps}
	Coupling of the impurity site (square) to the $t$-$J$ bulk sites
	(circles).}
\end{figure}

The solution of the model is completely analogous to that of the pure
$t$--$J$ model \cite{QISM_tJ}: The transfer matrix generating the
hamiltonian and the other conserved quantities is the trace of a product of
the local ${\cal L}$ operators chosen as ${\cal L}_{tJ}= (\lambda+i\Pi)/
(\lambda+i)$ for the regular sites ($\Pi$ is a graded permutation operator
acting on the tensor product of the auxiliary and the quantum space) and
${\cal L}_{34}\propto\lambda-i({\alpha \over 2}+1)+i\tilde{\cal L}$ on the
site associated to the impurity. Written as a matrix in the three
dimensional auxiliary space $\tilde{\cal L}$ reads
\[
 \tilde{\cal L}\!\!=\!\!
  \left( \begin{array}{ccc}
  X_2^{\downarrow \downarrow}+X_2^{00} &
  - X_2^{\downarrow \uparrow} &
   Q_{\uparrow}
\\
  - X_2^{\uparrow \downarrow} &
  X_2^{\uparrow \uparrow}+X_2^{00} &
  Q_{\downarrow} 	
\\
  Q^\dagger_{\uparrow} &	
  Q^\dagger_{\downarrow} &	
 \!\alpha+  X_2^{\uparrow \uparrow}+X_2^{\downarrow \downarrow}+2 X_2^{00}\!\!
\end{array} \right)\!\!.
\]
Here $Q_{\sigma}=\sqrt{\alpha+1}X_2^{0
\sigma}-\sigma\sqrt{\alpha}X_2^{-\sigma 2}$ with $\sigma=\pm$, where
$+$ ($-$) corresponds to $\uparrow$ ($\downarrow$) and where the
Hubbard projection operators are given by $X^{ab}=|a\rangle\langle b|$
with $a,b=\uparrow,\downarrow,2,0$. The hamiltonian is then given by the
logarithmic derivative of the transfer matrix at spectral parameter
$\lambda=0$. In general the form of this hamiltonian is rather complicated
and will be given elsewhere \cite{bef:up}. In any case the precise
form of the lattice (impurity) interactions is not essential as far as
low-energy properties are concerned: in the continuum limit only a
small number of terms with scaling dimensions smaller than two will
survive (taking the continuum limit and identifying the scaling
dimensions of the composite operators at the impurity site is rather
nontrivial though).
Physically the model describes an impurity with four allowed states
(spin-up/down electrons, empty/doubly occupied site) that couples to
two neighbouring $t$-$J$ sites and also modifies the interaction
between the $t$-$J$ sites (see Fig.~\ref{fig:im.ps}). In the
limiting cases $\alpha\to0(\infty)$ the hamiltonian simplifies
essentially: In the first case the impurity acts as a ordinary
$t$--$J$ site in the ground state below half filling, for
$\alpha\to\infty$ the impurity site is doubly occupied and induces a
twist in the boundary conditions of the host chain.

The eigenstates of the hamiltonian for $N_\uparrow$ ($N_\downarrow$)
electrons with spin $\uparrow$ ($\downarrow$) are constructed by means of
the nested algebraic Bethe Ansatz (NABA) leading to a system of algebraic
equations for the spectral parameters $\lambda_j$ ($j=1,\ldots,N_e=
N_\uparrow + N_\downarrow$) and $\lambda_\alpha^{(1)}$
($\alpha=1,\ldots,N_\downarrow$)
\begin{eqnarray}
 \left( \frac{\lambda_j-{i\over 2}}{\lambda_j+{i\over 2}} \right)^{L}
 \left( \frac{\lambda_j-{\alpha+1 \over 2}i}{\lambda_j+{\alpha+1 \over 2}i}
 \right) &=& \prod^{N_\downarrow}_{\alpha=1} \frac
 {\lambda_j-\lambda^{(1)}_\alpha-{i\over 2}}
 {\lambda_j-\lambda^{(1)}_\alpha+{i\over 2}}\ , \nonumber \\
 \prod^{N_e}_{j=1} \frac{\lambda^{(1)}_\alpha-\lambda_j+{i\over 2}}
 {\lambda^{(1)}_\alpha-\lambda_j -{i\over 2}} &=& -
 \prod^{N_\downarrow}_{\beta=1}
 \frac{\lambda^{(1)}_\alpha-\lambda^{(1)}_\beta+i}
 {\lambda^{(1)}_\alpha-\lambda^{(1)}_\beta-i}\ .
\nonumber
\end{eqnarray}
The corresponding eigenvalues of the hamiltonian in the grand canonical
ensemble are $E= -\mu N_e -(H/2)(N_\uparrow-N_\downarrow)
+\sum_{j=1}^{N_e}{1/( \lambda_j^2+{1\over 4})}$.

\begin{figure}
\noindent
\epsfxsize=0.4\textwidth
\epsfbox{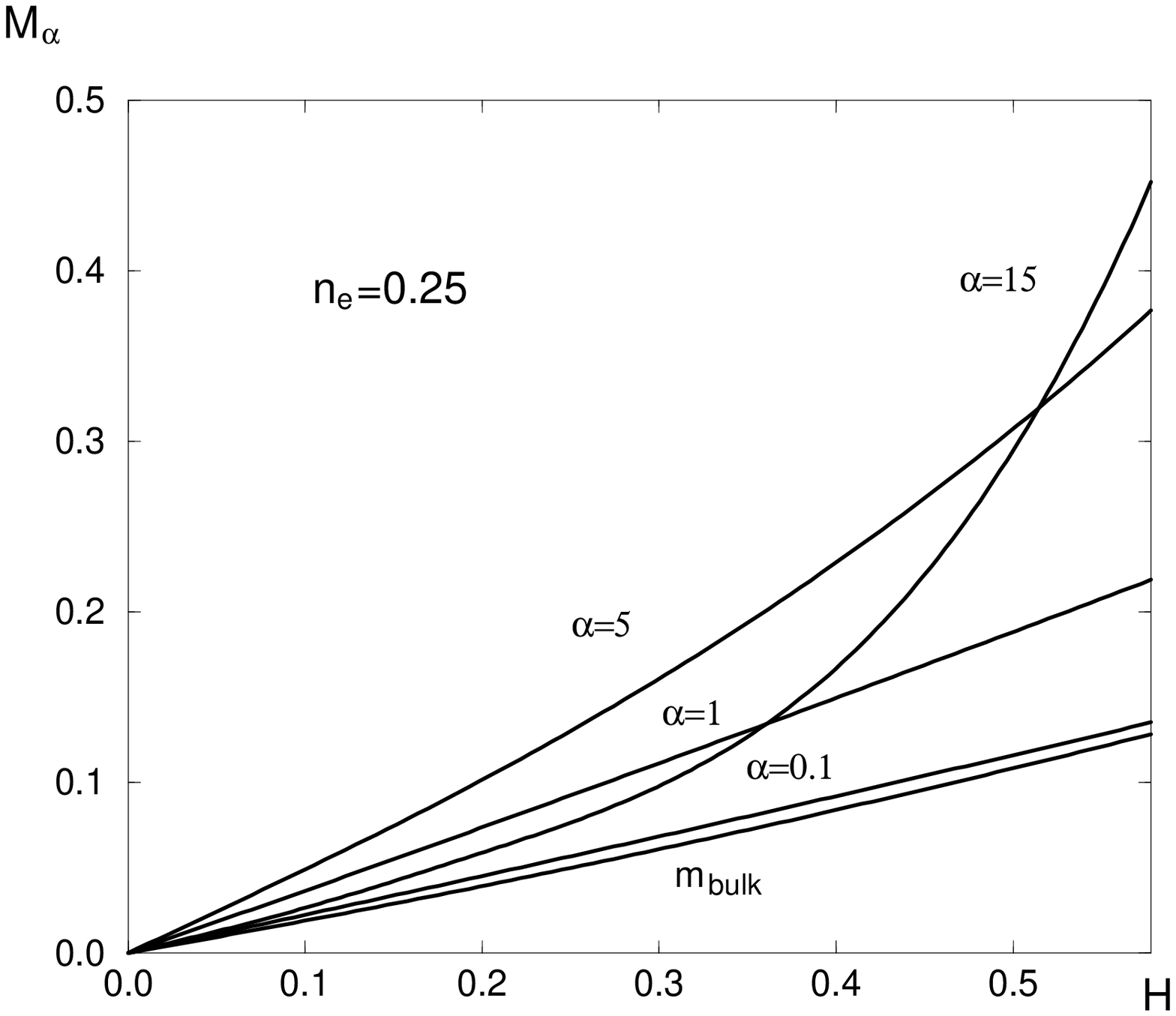}\\
\epsfxsize=0.4\textwidth
\epsfbox{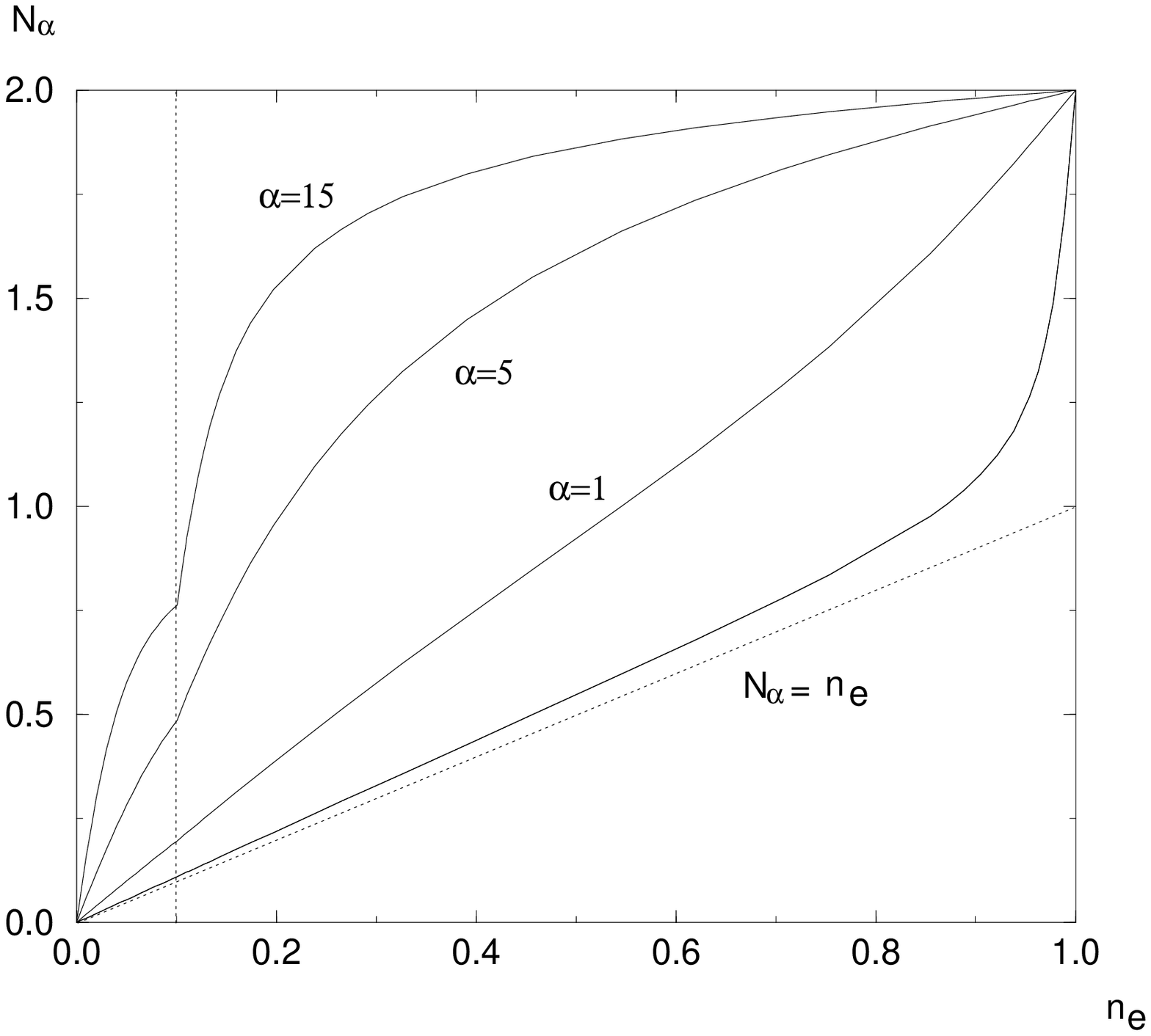}
\caption{\label{fig:gs_prop}
	(a) Impurity magnetization as a function of magnetic field for band
	filling $n_e=0.25$ and several values of $\alpha$. (b) Number of
	electrons located at the impurity as a function of the bulk
	electron--density for fixed magnetic field $H=0.1$ and several
	values of $\alpha$.  The dotted line denotes the critical
	electron--density $n_c$ below which the ground state is
	ferromagnetic.}
\end{figure}

%
%
The configuration of spectral parameters leading to the lowest energy state
for given chemical potential and magnetic field are found in complete
analogy to the pure $t$--$J$ chain: the ground state for finite $H$ is
described by two filled Fermi seas of $\lambda$-$\lambda^{(1)}$-``strings''
$\lambda_\pm=\lambda^{(1)} \pm \frac{i}{2}$ with real $\lambda^{(1)}$
associated with the holon excitations and real solutions $\lambda_j$
describing spin degrees of freedom. In the thermodynamic limit dressed
energies $\epsilon_c$ and $\epsilon_s$ can be associated with the
excitations of these objects. They are given in terms of coupled integral
equations which are identical to those found for the chain without
impurities \cite{schl:87}. In the resulting ground state energy the
impurity contribution can be identified from its $L$-dependence which
allows to compute the occupation, magnetization and susceptibilities of the
impurity site. Analytical results for these expectation values are
available only in limiting cases close to half filling and for densities
near or below the critical density $n_c$ related to the magnetic field by
$H=4\sin^2(\pi n_c/2)$ where the ground state is ferromagnetic and only
real $\lambda_j$ are present \cite{bef:up}.  For general values of band
filling and magnetic field the magnetization and particle number on the
impurity site can be determined by numerically solving a set of two coupled
integral equations (see Fig.~\ref{fig:gs_prop}): For $\alpha\to0$ the
impurity mimics the bulk behaviour as discussed above.  For large $\alpha$
the impurity occupation number is close to 1 (2) for $n_e<n_c$ ($>n_c$)
leading to an enhanced (reduced) magnetization at small (large) electron
densities.  The magnetic susceptibility of the impurity near half filling
is found to be $\sim \chi_{bulk}/\alpha$. We note that the magnetization
curves for sufficiently large $\alpha$ intersect the ones for small
$\alpha$.

%
%
In addition to the ground state properties the Bethe Ansatz allows to study
the finite temperature behaviour of the system. The thermodynamic Bethe
Ansatz equations for the system with impurity are the same as in the pure
case \cite{schl:87}. Again the impurity contribution can be isolated in the
free energy which can evaluated explicitly in the limit $T\to\infty$ and
for $H\gg T$ using Takahashi's method \cite{taka:74}. In the high
temperature limit we find
\begin{eqnarray}
  F_{\rm bulk}&=&-LT \ln\left(1+e^{\frac{\mu}{T}}2\cosh\frac{H}{2T}\right)
\nonumber\\
F_{\rm imp}&=& -\frac{2\alpha}{\alpha+2}-T
\ln\left(1+e^{\frac{2\mu}{T}}+2e^{\frac{\mu}{T}}\cosh\frac{H}{2T}\right)
\nonumber
\end{eqnarray}
giving the correct entropy in this limit. Note that the parameter $\alpha$
enters the leading term in this expansion in a trivial way only.

For low temperatures $T\ll H$ we can determine the phase diagram of the
system. Most interesting is the behaviour at half filling where the
impurity contribution to the specific heat is found to show a different
temperature dependence than the one from the bulk: for $2\mu>H>4$ the
system is ferromagnetic and the low T free energy is given by (we suppress
the contribution from the ground state energy)
\[
F_{\rm bulk} \approx -{L \over 2 \sqrt{\pi}}T^{3/2} e^{4-H \over T}, \qquad
F_{\rm imp}\approx  -{T} e^{H/2-\mu \over T}\ .
\]
For smaller magnetic fields the thermodynamic equilibrium state is not
ferromagnetically ordered, the bulk free energy is $F_{\rm bulk}
\approx-\pi L T^2/(6v_s)$ with the spinon velocity $v_s$ and the impurity
contribution is $F_{\rm imp}\approx -{2 \over \alpha}T^{3/2}$ up to factors
that cannot be caculated in closed form in general. Near $H=H_c$ this
factor becomes $\sqrt{3 \over 8}(4-H)^{-{3 \over 4}} \exp\left({1\over
T}(H/2-\mu+{2\over 3 \pi} (4-H)^{3/2})\right)$.

%
%
The effect of the impurities on the transport properties of the system can
be studied by calculating the spin- and charge stiffnesses from the finite
size corrections to the ground state energy of the model subject to twisted
boundary conditions \cite{Twist}. Following the analysis in \cite{bcfh:92}
we introduce twist angles $\phi_c$ and $\phi_s$ affecting charge and spin
degrees of freedom, respectively. The leading term in the resulting shift
of the ground state energy can then be written as $\Delta
E(\phi_c,\phi_s)=L^{-1} \phi_\alpha D_{\alpha\beta}(\alpha) \phi_\beta$
where charge (spin) stiffness are defined as
$D^{(\rho)}=(L/2)\partial^2_\phi \Delta E(\phi,0)=D_{cc}$
($D^{(\sigma)}=(L/2)\partial^2_\phi \Delta E(\phi,-2\phi)$). {}The analysis
of the finite-size corrections to the ground state energy yields the result
that for the case of a single impurity the stiffnesses are not modified to
leading order in $L^{-1}$. Hence, in spite of the presence of the impurity
we find an infinite dc-conductivity. This is completely different from the
situation in the ``weak-link'' type potential impurity discussed in
\cite{kafi:92,tsve:95a}: such a weak link drives the Luttinger liquid to a
strong coupling fixed point characterized by a vanishing conductivity. We
believe that the behaviour of the system considered here is related to its
integrability and the absence of backscattering at the impurity.

\begin{figure}
\noindent
\epsfxsize=0.4\textwidth
\epsfbox{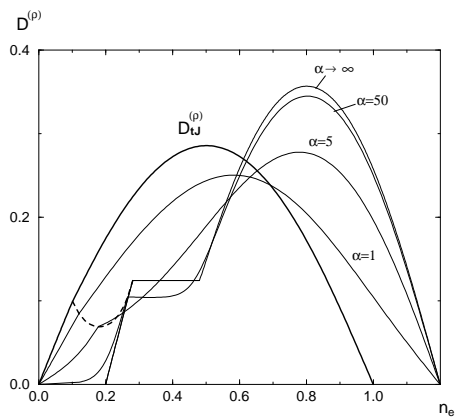}
\caption{\label{fig:stiff}%
	Charge--stiffness for a chain with $20$ percent impurities as a
	function of the electron density at $H=0.1$ for several values of
	$\alpha$. The dashed line denotes the stiffness at the critical
	electron density $n_c(\alpha)$. (For comparison we have included
	the stiffness for the pure $t$--$J$ chain $D_{tJ}^{(\rho)}$.)}
\end{figure}

The transport properties of the system do change if one considers a finite
density $n_i$ of impurity sites.  In this situation the band filling
can take values larger than 1 as the impurity sites allow for double
occupancies.  Comparing the charge stiffness to that of the pure $t$--$J$
case one observes a reduction for densities just above the critical
one. For larger band fillings the presence of the impurities leads to an
enhancement of the stiffness (see Fig.~\ref{fig:stiff}). This is easily
understood: $D^{(\rho)}$ vanishes at half filling in the $t$--$J$
chain. The impurities do allow double occupancies thereby enlarging the
phase space for the electrons which leads to an increase in the
stiffness. At large fillings the stiffness increases as a function of
$\alpha$ as the average occupation number of the impurity sites are close
to doubly occupied which makes the movement of electrons between the
$t$--$J$ sites easier. In particular, the ``absorption'' of particles by
the impurity sites for $\alpha=\infty$ leads to plateaus in the stiffness
for $n_e<n_i$ (where it vanishes) and for
$n_i+(1-n_i)n_c<n_e<2n_i+(1-n_i)n_c$. For other fillings the stiffness is
simply given by that of the $t$--$J$ model (up to a rescaling).  Similarly
the reduction of the spin stiffness due to the addition of impurities can
be understood.

%
%
Finally, to study the effects of the impurity on the excitations in the
model we have computed the phase shifts acquired by holons and spinons due
to scattering off the impurity in the case of a microscopic number of holes
in the half filled ground state at vanishing magnetic field.  The basic
ingredient for this calculation is the quantization condition for
factorized scattering of two particles with rapidities $\lambda_1$ and
$\lambda_2$ on a ring of length $N$, namely $\exp(iNk(\lambda_1))
S(\lambda_1-\lambda_2) \exp(\psi(\lambda_1))=1$ where $k(\lambda)$ is the
physical momentum in the infinite periodic system, $S(\lambda)$ is the bulk
scattering matrix for scattering of particles 1 and 2 and $\psi(\lambda_1)$
is the phase shift acquired when scattering off the impurity (note that
this incorporates the fact that there is no backscattering at the
impurity).  Using the known result for $S$ \cite{babo:91} one extracts the
impurity phase shifts using the method of \cite{kore:79,al}: both the
spinon and holon impurity phase shifts are proportional to
$\exp(-ik)$, where $k$ denotes the physical momentum of spinons/holons
in the $t$-$J$ model {\sl without impurity}. This reflects the fact
that the impurity essentially decouples from the chain at half filling
leading to a chain of $N-1$ sites.  In addition the holons pick up a
phase shift $(2i\lambda-\alpha)/(2i\lambda+i\alpha)$ 
due to the fact that the impurity site is charged. The pole at
$\lambda=i\alpha/2$ corresponds to an impurity bound state for
$\alpha<2$. 

To summarize, we have studied the effects of the addition of integrable
impurities to the supersymmetric $t$--$J$ model on certain zero and
finite temperature properties of the system.  The properties of the
impurity, which couples to both spin and charge degrees of freedom,
can be tuned by adjusting a continous parameter $\alpha$.
Compared to the ``weak-link'' type impurities investigated by Kane and
Fisher \cite{kafi:92} it appears to be very special in that its
dc-conductivity is unchanged by the addition of a single impurity.  We
have argued that this is due to the absence of backscattering terms on
the level of the dressed excitations (holons and spinons).  Although
the verification by explicit construction of the continuum limit for
this system appears difficult one may speculate that similar to the
case of a Kondo impurity in a Luttinger liquid \cite{FrojdhJ} a
backscattering term would drive the system to a new fixed point. Hence
the present model can be interpreted as an unstable fixed point from a
renormalization group point of view.

This work has been supported in parts by the Deutsche
For\-schungs\-gemeinschaft under Grant No.\ Fr~737/2--2.

\setlength{\baselineskip}{13pt}


\end{multicols}
\end{document}